\documentclass[copyright]{eptcs}
\providecommand{\event}{T. Neary, D. Woods, A.K. Seda and N. Murphy (Eds.):\\ The Complexity of Simple Programs 2008.}
\providecommand{\volume}{1}
\providecommand{\anno}{2009}
\providecommand{\firstpage}{16}
\usepackage{breakurl}

\usepackage{latexsym,url}             
\newtheorem{thm}{Theorem}
\newtheorem{cor}[thm]{Corollary}

\newcommand{\N}{{\bf N}}
\newenvironment{proof}{\trivlist \item[\hskip \labelsep{\bf Proof.}]}
{{\hfill$\Box$}\endtrivlist}

\title{Simplicity via Provability for Universal Prefix-free {T}uring Machines}
\author{Cristian S. Calude
    \email{\url{www.cs.auckland.ac.nz/~cristian}}
    \institute{ Department of Computer Science\\
    University of Auckland\\
    Private Bag 92019, Auckland, New Zealand}
    }

\def\titlerunning{Simplicity via Provability for Universal Prefix-free {T}uring Machines}
\def\authorrunning{C. S. Calude}

\begin{document}
\maketitle

\begin{abstract}
Universality is  one of the most important ideas in computability theory.
There are various criteria of simplicity for universal Turing machines. Probably
the most popular one is to count the number of states/symbols.
This criterion is more complex than it may appear at a first glance.
In this note we review recent results in Algorithmic Information Theory and propose three new criteria of simplicity for universal prefix-free Turing machines. These criteria refer to the possibility of proving various natural properties of such a machine (its universality, for example) in a formal theory,  PA or ZFC.   In all cases some, but not all, machines are simple. 
\end{abstract}

\section{The smallest universal Turing machine}

Roughly speaking, a universal   Turing machine is a Turing machine capable of simulating any other  Turing machine.  In Turing's words:
\begin{quote}\it
It can be shown that a single special machine of that type can be made to do the work of all. It could in fact be made to work as a model of any other machine. The special machine may be called the universal machine.
\end{quote}

The first universal Turing machine was constructed by Turing  \cite{AT,ATc}.
Shannon \cite{CS} studied the problem  of finding the smallest possible universal Turing machine and showed that two symbols were sufficient, if enough states can be used. 
He also proved that  ``it is possible to exchange symbols for states and vice versa (within certain limits) without much change in the product.'' Notable universal Turing machines  include the machines constructed by
 Minsky    (7-state 4-symbol) \cite{MM},   Rogozhin (4-state 6-symbol) \cite{YR}, 
 Neary--Woods (5-state 5-symbol) \cite{NW}.
Herken's book \cite{Herken} celebrates the first 50 years of universality.
 
 Weak forms of universality were proved by
 Watanabe	(4-state 5-symbol) \cite{Watanabe}, Cook \cite{Cook} for
 Wolfram's 2-state 5-symbol machine \cite{SW}, Neary--Woods \cite{NW1}, and Smith \cite{Smith} for Wolfram's  2-state 3-symbol  machine.\footnote{The critique  by Pratt \cite{VP1,VP2}, the response in \cite{NKSForum} and the forthcoming paper by Margenstern \cite{Maurice} show the subtlety of the notion of universality.}

\section{Universal prefix-free Turing machines}

 A prefix-free Turing machine, shortly, machine, is a Turing machine whose domain is a prefix-free set. In what follows we will be concerned only with machines working on the binary alphabet $\{0,1\}$. A {\it universal machine} $U$ is a machine such that for every other machine $C$ there exists a constant $c$ (which depends upon $U$ and $C$) such that for every program $x$ there exists a program $x'$ with $|x'| \le |x|+c$ such that $U(x')=C(x)$. Universal machines can be effectively constructed. For example, given a  computable enumeration of all machines $(C_{i})_{i}$,
 the machine $U$ defined by  $U(0^{i}1x) = C_{i}(x)$ is universal.\footnote{See more in \cite{Ca}. The above universal machine, called {\em prefix-universal} because universality is obtained 
by adjunction, is quite particular. There are universal machines which are not prefix-universal. } The domains of universal machines have interesting computational and coding properties, cf.
 \cite{CL,CNSS}.

 \section{Peano arithmetic and Zermelo--Fraenkel set theory}
 
 By $\mathcal{L}_{A}$ we denote the first-order language   of arithmetic whose non-logical  symbols consist of  the constant symbols 0 and 1, the binary relation symbol $<$ and  two binary function symbols $+$ (addition) and $\cdot$ (multiplication). Peano arithmetic, PA, is  the first-order theory \cite{RK} given by a set of 15 axioms defining discretely ordered rings,  together with induction axioms for each formula $\varphi (x,y_{1},\ldots ,y_{n})$ in $\mathcal{L}_{A}$:
\[\forall \overline{y}(\varphi(0,\overline{y}) \wedge \forall x(\varphi(x,\overline{y}) \rightarrow \varphi(x+1,\overline{y})) \rightarrow \forall x (\varphi(x,\overline{y})).\]
By PA $\vdash \theta$ we mean ``there is a proof in PA for $\theta$''.  

PA is a first-order theory of arithmetic powerful enough to prove many important
results in computability and complexity theories. For example, there are total computable functions for which PA cannot prove their totality, but PA can prove the totality of 
every primitive recursive function (and also of Ackermann total computable, non-primitive recursive function), see \cite{RK}.

Zermelo--Fraenkel set theory with the axiom of choice, ZFC, is  the standard  
one-sorted first-order theory of sets;  it is considered the most common foundation of mathematics. In ZFC
set membership  is a primitive relation.  By ZFC $\vdash \theta$ we mean ``there is a proof in ZFC for $\theta$''.  

Our metatheory is ZFC. We fix a (relative) interpretation of PA in ZFC according to which each formula of $\mathcal{L}_{A}$ has a translation into a formula of ZFC. By abuse of language we shall use the phrase ``sentence of arithmetic''  to mean a sentence (a formula with no free variables) of ZFC  that is the translation of some formula of PA.

\section{Rudiments of Algorithmic Information Theory}
 The set of bit strings is denoted by $\Sigma^{*}$. If $s$ is a bit string then $|s|$ denotes the length of $s$. 
 All reals will be in the unit interval.  A computably enumerable (shortly, c.e.)  real number $\alpha$ 
is given by an increasing computable sequence of rationals converging to $\alpha$.
Equivalently, a c.e.\ real  $\alpha$ is the limit of 
an increasing primitive recursive sequence  of rationals. 
We will blur the distinction between the real $\alpha$ and the infinite base-two expansion of
$\alpha$, i.e.\  the infinite bit sequence $\alpha_{1} \alpha_{2}\cdots \alpha_{n}\cdots $ ($\alpha_{n}\in\{0,1\})$ such that $ \alpha = 0.\alpha_{1} \alpha_{2}\cdots \alpha_{n}\cdots $.  By
$\alpha(n)$ we denote the  string of length $n$, $\alpha_{1} \alpha_{2}\cdots \alpha_{n}$.

One of the major problems in algorithmic information theory is to define and study
(algorithmically) random reals. To this aim one can use the prefix-complexity or
constructive measure theory; remarkably,  the class of ``random reals'' obtained with different approaches remains the same.  

In what follows we will adopt the complexity-theoretic approach. Fix a universal machine $U$. The prefix-complexity induced by $U$  is the function $H_{U}: \Sigma^{*} \rightarrow \N$ ($\N$ is the set of natural numbers) defined by the formula: $H_{U}(x) = \min\{|p| \; : \, U(p)=x\}$. One can prove that this complexity is optimal up to an additive constant in the class of all prefix-complexities $\{H_{C} \,:\,
C \mbox{ is a machine}\}$. 

  A c.e.\ real $\alpha$ is {\em Chaitin-random} if there exists  a constant $c$ such that for all $n\ge 1$, $H_{U}(\alpha (n))\ge n-c$.  The above definition is invariant with respect to  $U$. Every Chaitin-random real is non-computable, but the converse is not true. Chaitin-random reals abound: they have (constructive) Lebesgue measure one, cf. \cite{Ca}.
  
  The standard example of c.e.\  Chaitin-random real is the 
halting probability of a universal machine $U$(Chaitin's Omega number):\footnote{$U(x)<\infty$ means ``$U$ is defined on $x$''.} $$\Omega_{U}
= \sum_{U(x) <\infty} \, 2^{-|x|}.$$ 

Each Omega number encodes information about halting programs in the most compact way. For example, 
the answers to the following $2^{n+1}-1$ questions ``Does $U(x)$ halt?'', for all programs $|x|\le n$,  is  encoded in the first $n$ digits of $\Omega_{U}$---an exponential rate of compression. Is this important? For example, to solve the Riemann hypothesis
one needs to calculate the first 7,780 bits of a natural Omega number \cite{CEM}.

The following result characterises the class of c.e.\  Chaitin-random reals:

 {\thm {\bf \cite{chaitin75,CHKW98stacs,slaman}}  \label{repth} The set of 
c.e.\  Chaitin-random reals coincides with the set of all halting probabilities of all universal  machines.}

\medskip

C.e.\ random reals
have been intensively studied in recent years, with many results summarised in \cite{Ca,DH,Nies}.

\section{Universal machines simple for PA}

We start with the simple question: Can PA certify the universality of a universal machine?

\medskip

A universal machine $U$ is called {\it simple for} PA if PA $\vdash ``U$ is universal'',
i.e. 
PA can prove that a universal $U$, given by its full description, is indeed universal.
For illustration, the results in this section will include full proofs.

\medskip

As one might expect, there exist universal machines simple for PA:

{\thm {\bf \cite{CH}}  One can effectively construct a universal machine  which is simple for {\rm PA}.}

\begin{proof} The set of all  machines PA can prove to be   prefix-free is c.e., so if 
$(C_i)_{i}$ is a computable enumeration of provably prefix-free machines, then the machine $U_{0}$ defined by $U_{0}(0^{i}1x) = C_i(x)$ has the property specified in the theorem: {\rm PA} $\vdash ``U_{0}$ is universal''.\end{proof}

\medskip

However,  not all universal machines are simple:

{\thm {\bf \cite{CH}} One can effectively construct a universal machine  which is not simple for
{\rm PA}.}

\begin{proof}
Let $(f_{i})_{i}$ be a c.e.\ enumeration of all primitive recursive functions $f_{i}: \N \rightarrow
\Sigma^{*}$ and $(C_{i})_{i}$ a c.e.\ enumeration of all prefix-free machines. Fix a universal  prefix-free machine $U$ and consider the computable function $g: \N \rightarrow \N$ defined by:

\[ C_{g(i)}(x) = \left\{ \begin{array}{ll}
 U(x), & \mbox{\rm if for some $j>0,  \#\{f_{i}(1), f_{i}(2), \ldots ,f_{i}(j)\}> |x|$}, \\
\infty, & \mbox{\rm otherwise} \,.
  \end{array} \right.\]

For every $i$, $C_{g(i)}$ is a  prefix-free universal machine iff $f_{i}(\N)$ is infinite (if $f_{i}(\N)$ is finite, then so is  $C_{g(i)}$). Since the set
of all indices of primitive recursive functions with infinite range is not c.e.\
it follows that PA cannot prove that for some $i, C_{g(i)}$ is   universal.
\end{proof}

\medskip

Both results above are true for plain universal machines too. The above proofs work for plain universal machines, but a simpler proof can be given for the negative result.

\section{Universal machines simple for  {\rm \bf ZFC}}

Assume that the binary expansion of $\Omega_{U}$ is $0.\omega_{1}\omega_{2}\cdots $. For each digit $\omega_{i}$ we can consider two
arithmetic sentences in ZFC, ``$\omega_{i}=0$'', ``$\omega_{i}=1$''. How many sentences of the above   type   can ZFC prove?

{\thm {\bf \cite{chaitin75}} Assume that  {\rm ZFC} is arithmetically
sound (that is,  each sentence of arithmetic proved by  {\rm ZFC} is true).
Then, for every   universal   machine $U$,  {\rm ZFC} can determine
the value of only finitely many bits of the binary expansion of $\Omega_U$,  and one can
calculate a bound on the number of bits of
$\Omega_U$ which
 {\rm ZFC} can determine.\footnote{This means that ZFC can prove only finitely many sentences of the form 
``$\omega_{i}=0$'', ``$\omega_{i}=1$'' and one can calculate a natural $N$ such that no sentence of the above type with $i\ge N$ can be proved in ZFC.}}

\medskip

Actually, we can precisely describe the``moment'' ZFC fails to prove any bit of 
$\Omega_U$:

{\thm {\bf \cite{Incompl} }  \label{cris} Assume that {\rm ZFC} is arithmetically sound. Let $i \ge 1$ and consider the c.e.~random real 
\[ \alpha = 0.\alpha_1 \ldots \alpha_{i-1}\alpha_i \alpha_{i+1}\ldots, \mbox{   where  
}  
\alpha_1 =
\ldots
=\alpha_{i-1} = 1, \alpha_{i}=0.\] 
Then, we can effectively construct  a universal   machine $U$ (depending upon
 {\rm ZFC} and $\alpha$)  such that  {\rm PA} proves the universality of $U$,
 {\rm ZFC} can  determine at most $i$ initial bits of $\Omega_U$ and $\alpha = \Omega_U$.}

\medskip

In other words, the moment the first 0 appears (and this is always the case because
$\alpha$ is random) ZFC cannot prove anything about the values of the remaining bits.

By taking $\alpha < 1/2$ we get Solovay's most ``opaque'' universal   machine:\footnote{Theorem~\ref{solovay} was obtained before Theorem~\ref{cris}.}

\medskip

{\thm {\bf \cite{solovay2k}} \label{solovay} One can effectively construct a  universal   machine $U$
such  that   {\rm ZFC} (if arithmetically sound) cannot determine any bit of $\Omega_U$. }

\medskip

 We say that a universal machine is {\it $n$--simple for} ZFC if ZFC can prove at most $n$ digits of the binary expansion of $\Omega_{U}$. In view of Theorem~\ref{cris}, for every $n\ge 1$ there exists a  universal machine which is {\it $n$--simple for} ZFC.  By Theorem~\ref{solovay} there exists a universal machine which is not 1--simple for ZFC.

\section{Universal machines {\rm \bf PA}--simple for randomness}

We first express Chaitin randomness in PA. 
A c.e.\ real $\alpha$ is {\em provably Chaitin-random} if there exists a universal   machine simple for  {\rm PA} and a constant $c$ such 
that  {\rm PA} $\vdash ``\forall n (H_{U}(\alpha (n))\ge n-c)$''.

In this context it is natural to ask the question: Which universal machines $U$
``reveal'' to PA that $\Omega_{U}$ is Chaitin-random?

{\thm {\bf \cite{CH}} \label{Chaitinrevisited}The halting probability of a  universal   machine  simple for {\rm PA} is provably Chaitin-random}.

\medskip

In fact, Theorem~\ref{repth} can be proved in PA:

{\thm {\bf \cite{CH}} The set of 
c.e.\  provably Chaitin-random reals coincides with the set of all halting probabilities of all  universal  machines simple for {\rm PA}.}

\medskip

Based on Theorem~\ref{Chaitinrevisited} we define another (seemingly more general) notion of randomness in PA.  A c.e.\ real is {\em provably-random} (in PA) if there is a universal machine simple for PA and  PA $\vdash ``\Omega_{U}=\alpha$''.

\medskip

{\thm {\bf \cite{CH}} \label{ceprovrandc}  A c.e.\ real is provably-random iff it is  provably Chaitin-random.}

\medskip

In contrast with the case of finite random strings where ZFC (hence PA) cannot prove the randomness of more than finitely many strings, for c.e.\ reals we have:

\medskip

{\thm {\bf \cite{CH}} \label{ceprovrand}   Every c.e.\  random real is provably-random.}

\medskip

We say that  a universal machine $U$ is PA--{\it simple for randomness} if
{\rm PA} $\vdash ``\Omega_{U}$ is random.'' In view of the Theorem~\ref{ceprovrand} we get:

{\cor For every c.e.\  random real $\alpha$ there exists a {\rm PA}--simple for randomness universal machine $U_{0}$  such that $\alpha = \Omega_{U_{0}}$.}

 \medskip However,

{\thm There exists a universal machine which is not  {\rm PA}--simple for  randomness.} 

\section{Conclusions}
We have used some recent results in Algorithmic Information Theory to introduce three new criteria of simplicity for universal machines based on their
``openness'' in revealing information to a formal system, PA or ZFC. The type of encoding is essential for these criteria. This  point of view might be useful in other contexts, specifically in automatic theorem proving. It would be interesting to ``actually construct'' the universal machines discussed in this paper.

\section*{Acknowledgement} 
I thank D. Woods whose invitation to CSP08 stimulated these thoughts and H. Zenil who helped me with recent references. 
I am indebted to the anonymous referees for their comments which substantially improved the presentation. 

\bibliographystyle{eptcs}

\begin{thebibliography}{}
\providecommand{\bibitemstart}[1]{\bibitem{#1}}
\providecommand{\bibitemend}{}
\providecommand{\bibliographystart}{}
\providecommand{\bibliographyend}{}
\providecommand{\url}[1]{\texttt{#1}}
\providecommand{\urlprefix}{Available at }
\providecommand{\bibinfo}[2]{#2}
\bibliographystart

\bibliographyend
\end{thebibliography}


\begin{thebibliography}{99}
 \bibitem{Ca} C. S. Calude. {\em Information and Randomness. An Algorithmic
    Perspective}, 2nd Edition, Revised and Extended, Springer Verlag, Berlin,
  2002.
  \bibitem{Incompl} C.~S.~Calude. Chaitin $\Omega$ numbers, Solovay machines
and incompleteness, 
 {\em
Theoret. Comput. Sci.}  284 (2002), 269--277.

\bibitem{CEM}  C. S. Calude, Elena Calude, M. J. Dinneen. A new measure of the difficulty  of  problems,  {\em Journal for Multiple-Valued Logic and Soft Computing} 12 (2006), 285--307.

  \bibitem{CH}  C. S. Calude, N. J. Hay. Every Computably Enumerable Random Real Is Provably Computably Enumerable Random,
{\em CDMTCS Research Report} 328, 2008, 29 pp.

   \bibitem{CHKW98stacs}
C.~S.~Calude, P.~Hertling, B.~Khoussainov, and Y.~Wang.
Recursively enumerable reals and {C}haitin $\Omega$ numbers,
in: M.~Morvan, C.~Meinel, D.~Krob \ (eds.),
{\em Proceedings of the 15th Symposium on
Theoretical Aspects of Computer Science (Paris)},
Springer--Verlag, Berlin, 1998, 596--606. Full paper  in {\em Theoret.
Comput. Sci.}  255 (2001), 125--149.

\bibitem{CNSS} C.  S. Calude, A. Nies, L. Staiger,  F. Stephan. 
Universal recursively enumerable sets of strings, in M. Ito, M. Toyama (eds.). 
{\it Developments in Language Theory (DLT'08)}, Lectures
Notes in Comput. Sci.  5257, Springer-Verlag, Berlin, 2008,  170--182.

\bibitem{CL}  C. S. Calude,  L. Staiger. On universal computably enumerable prefix codes,  {\em Mathematical Structures in Computer Science} 19 (2009), 45--57.


  \bibitem{chaitin75}
G. J. Chaitin. A theory of program size formally identical to
information theory, {\em J. Assoc. Comput. Mach.} 22 (1975), 329--340.

\bibitem{Cook} M. Cook. Universality in elementary cellular automata,
{\em Complex Systems} 15(1) (2004), 1--40. 

 \bibitem{DH} R. Downey, D. Hirschfeldt. {\em Algorithmic Randomness
    and Complexity}, Springer, Heidelberg,  2008.


\bibitem{Herken} R. Herken. {\em The Universal Turing Machine: A Half-Century Survey}, 		
Oxford University Press, Oxford, 1992.

   \bibitem{RK} R. Kaye. {\em Models of Peano Arithmetic}, Oxford Press, Oxford, 1991.
   \bibitem{slaman} 
 A. Ku\v{c}era, T.~A. Slaman. Randomness and recursive enumerability, {\em SIAM J. Comput.}
31, 1 (2001), 199-211.

\bibitem{Maurice} M. Margenstern. Turing machines with two letters and two 
states, {\em Complex Systems}, to appear.

\bibitem{MM} M. Minsky.  Size and structure of universal Turing machines using Tag systems, in {\em Recursive Function Theory, Proc. Symp. Pure Mathematics},  AMS, Providence RI, 5, 1962,  229--238.

\bibitem{NW1} D. Woods, T. Neary. Small semi-weakly universal Turing machines,
{\em Fundamenta Informaticae}, 91(1):179-195, 2009.


\bibitem{NW} T. Neary, D.  Woods. Four small universal Turing machines,
{\em Fundamenta Informaticae}, 91(1):123-144, 2009.
  
\bibitem{Nies} A. Nies. {\em Computability and Randomness}.
Oxford University Press, 2009.

\bibitem{NKSForum} NKS Forum, \url{http://forum.wolframscience.com/showthread.php?s=&threadid=1472}.

\bibitem{VP1} V. Pratt. Simple Turing machines, universality, encodings, etc.,
\url{http://cs.nyu.edu/pipermail/fom/2007-October/012156.html}.
\bibitem{VP2} V. Pratt. Definition of universal Turing machine, \url{http://cs.nyu.edu/pipermail/fom/2007-October/012148.html}.



\bibitem{YR} Y. Rogozhin. A universal Turing machine with 22 states and 2 symbols, {\em Romanian Journal of Information Science and Technology} 1 (3) (1998), 259--265.

\bibitem{CS} C. Shannon. A universal Turing machine with two internal states, in  {\em Automata Studies, Princeton},  Princeton University Press, NJ, 1956, 157--165.

\bibitem{Smith} A. Smith. WolframÕs 2,3 Turing machine is universal, {\em Complex Systems}, 
to appear. 

\bibitem{solovay2k}
R. M. Solovay.  A version of $\Omega$ for which $ZFC$ can not predict a
single bit, in C. S. Calude, G. P\u{a}un (eds.). {\em Finite Versus
Infinite. Contributions to an Eternal Dilemma}, 
Springer-Verlag, London, 2000, 323--334.

\bibitem{AT} A. Turing. On computable numbers, with an application to the Entscheidungsproblem, {\em Proceedings of the London Mathematical Society} 42 (2) (1936), 230--265.


\bibitem{ATc} A. Turing. On computable numbers, with an application to the Entscheidungsproblem: A correction,{\em Proceedings of the London Mathematical Society} 2, 43 (1937), 544--546.

\bibitem{Watanabe}  S. Watanabe. 4-symbol 5-state universal Turing machine,
{\em Information Processing Society of Japan Magazine} 13 (9) (1972), 588--592.

\bibitem{SW} S. Wolfram. {\em A New Kind of Science}, Wolfram Research, 2002,  706--714.

\bibitem{Wolfram} Wolfram 2,3 Turing Machine,\\ \url{http://demonstrations.wolfram.com/TheWolfram23TuringMachine/}.


\end{thebibliography}

\end{document}